\documentclass[aps,prd,amsmath,amssymb,preprint]{revtex4-1}

\usepackage{array} 
\usepackage{amssymb}
\usepackage{graphics,graphpap}
\usepackage{graphicx}
\usepackage{color}
\usepackage{graphicx}
\usepackage{dcolumn}
\usepackage{epsfig}
\usepackage{epstopdf}
\usepackage{bm}
\usepackage{amsmath}
\usepackage{amsfonts}
\usepackage{textcomp}
\usepackage{setspace}
\usepackage{natbib}
\usepackage{subfigure}
\usepackage{caption}

\newcommand{\beq}{\begin{eqnarray}}
\newcommand{\eeq}{\end{eqnarray}}

\newcommand{\im}{{\rm i}}
\newcommand{\dd}[2]{\frac{{\rm d}#1}{{\rm d}#2}}
\newcommand{\bmp}{\noindent\begin{minipage}{16cm}}
\newcommand{\emp}{\end{minipage}\vskip 7mm} 

\def\drawbox#1#2{\hrule height#2pt
        \hbox{\vrule width#2pt height#1pt \kern#1pt
              \vrule width#2pt}
              \hrule height#2pt}

\def\Asym#1#2{\vcenter{\vbox{\drawbox{#1}{#2}
              \kern-#2pt 
              \drawbox{#1}{#2}}}}

\def\simge{\mathrel{%
   \rlap{\raise 0.511ex \hbox{$>$}}{\lower 0.511ex \hbox{$\sim$}}}}

\def\simle{\mathrel{
   \rlap{\raise 0.511ex \hbox{$<$}}{\lower 0.511ex \hbox{$\sim$}}}}

\def\s#1{\setbox0=\hbox{$#1$}%
\rlap{\ifdim\wd0>.7em\kern.22\wd0\else\kern.1\wd0\fi /}#1}

\newcommand{\GeV}{\,\mbox{GeV}}

\newcommand{\ul}{{\rm L}}
\newcommand{\ur}{{\rm R}}
\newcommand{\hc}{{\rm h.c.}}

\newcommand{\ie}{i.e.}
\newcommand{\eg}{e.g.}

\newcommand{\todo}[1]{(\textbf{TODO:} #1)}
\renewcommand{\todo}[1]{}

\def\slc#1{\setbox0=\hbox{$#1$}           
    \dimen0=\wd0                                 
    \setbox1=\hbox{/} \dimen1=\wd1               
    \ifdim\dimen0>\dimen1                        
       \rlap{\hbox to \dimen0{\hfil/\hfil}}      
       #1                                        
    \else                                        
       \rlap{\hbox to \dimen1{\hfil$#1$\hfil}}   
       /                                         
    \fi}

\begin{document}

\title{Running of Neutrino Parameters and the Higgs Self-Coupling in a Six-Dimensional UED Model}

\date{\today}

\author{Tommy Ohlsson}
\email{tohlsson@kth.se}

\author{Stella Riad}
\email{sriad@kth.se}

\affiliation{Department of Theoretical Physics, School of Engineering Sciences, KTH Royal Institute of Technology -- AlbaNova University Center, Roslagstullsbacken 21, 106 91 Stockholm, Sweden}
\begin{scriptsize}
{\tiny}
\end{scriptsize}
\begin{abstract}

We investigate a six-dimensional universal extra-dimensional model in the extension of an effective neutrino mass operator. We derive the $\beta$-functions and renormalization group equations for the Yukawa couplings, the Higgs self-coupling, and the effective neutrino mass operator in this model. Especially, we focus on the renormalization group running of physical parameters such as the Higgs self-coupling and the leptonic mixing angles. The recent measurements of the Higgs boson mass by the ATLAS and CMS Collaborations at the LHC as well as the current three-flavor global fits of neutrino oscillation data have been taken into account. We set a bound on the six-dimensional model, using the vacuum stability criterion, that allows five Kaluza--Klein modes only, which leads to a strong limit on the cutoff scale. Furthermore, we find that the leptonic mixing angle $\theta_{12}$ shows the most sizeable running, and that the running of the angles $\theta_{13}$ and $\theta_{23}$ are negligible. Finally, it turns out that the findings in this six-dimensional model are comparable with what is achieved in the corresponding five-dimensional model, but the cutoff scale is significantly smaller, which means that it could be detectable in a closer future.
\end{abstract}

\maketitle

\section{\label{sec:intro}Introduction}
The primary goal of the Large Hadron Collider (LHC) at CERN in Geneva, Switzerland is to confirm or reject the existence of a Higgs boson. Since both the ATLAS and CMS Collaborations have observed a Higgs boson-like particle with mass of around 126 GeV at a significance of $5\sigma$ earlier this year, the goal seems to have been reached \cite{:2012gk,:2012gu}. Apart from searches for the Higgs boson, there is an ongoing hunt for signals of physics beyond the Standard Model (SM) at the LHC, and hopefully these experiments will shed some light on physics at the TeV scale.

One popular way to extend the SM is by increasing the number of space-time dimensions and it turns out that it is only possible to increase the number of spatial dimensions. The idea of extra dimensions originates from Theodor Kaluza and Oskar Klein in the 1920s \cite{Kaluza:1921tu,Klein:1926tv}. However, these theories were forgotten for decades until the 1980s, when extra-dimensional models had a revival at the birth of string theory. At the beginning of the new millennium, several extra-dimensional models had been established, that eventually can be detected in future high-energy collider experiments. These models have some nice properties such as providing a good dark matter candidate as well as the answers to the hierarchy problem and the problem of proton stability \cite{ArkaniHamed:2000hv,Hashimoto:2003ve,Hashimoto:2004xz,Servant:2002aq,Appelquist:2001mj,Cheng:2002ej}.

In the present work, we study a model with universal extra dimensions (UEDs), where the SM fields are allowed to propagate through the extra, spatial dimensions. This will give rise to an infinite tower of heavy Kaluza--Klein (KK) modes corresponding to each SM particle. The first UED model, which was introduced in 2000 by Appelquist, Dobrescu, and Cheng, has one extra dimension compactified on a circle \cite{Appelquist:2000nn}. In this Letter, we will study a six-dimensional UED model \cite{Burdman:2005sr}.

The KK modes interact with the SM particles and, given that the KK modes are at the TeV scale, they will therefore affect the renormalization group (RG) running of physical parameters in the UED model. The logarithmic behavior of the RG running in the SM will change to a power-law behavior in the extra-dimensional model \cite{Dienes:1998vg}. RG running in extra-dimensional models has been extensively studied, see \eg ~Refs.~\cite{Bhattacharyya:2006ym,Blennow:2011tb,Blennow:2011mp,Cornell:2011ge}.

One of a few rare phenomena, that requires beyond SM physics and actually has been established, is that of neutrino oscillations, which indicate that neutrinos are massive and lepton flavors are mixed. In contrast to other leptonic masses in the SM, the neutrino masses cannot be generated through the Higgs mechanism. It is thus impossible to describe the neutrino masses with the particle content of the SM, and some kind of new physics is required. In this work, we extend the six-dimensional UED model with a dimension-five Weinberg operator, which generates the neutrino masses \cite{Weinberg:1980bf}.

The work is organized as follows. In Sec.~\ref{sec:UED}, we discuss the six-dimensional UED model and present the renormalization group equation (RGE) for the effective neutrino mass operator. Then, in Sec.~\ref{sec:RGrunning}, we discuss RG running in extra-dimensions in general. Next, in Sec.~\ref{sec:RGHiggs}, we use the vacuum stability criterion, and the recent Higgs mass results, of the Higgs self-coupling constant in order to put constraints on the six-dimensional UED model. Furthermore, in Sec.~\ref{sec:RGMixing}, we discuss the running of the leptonic mixing angles in the six-dimensional UED model. Finally, in Sec.~\ref{sec:summary}, we summarize and present our conclusions. In addition, in Appendix \ref{sec:app}, we also provide the $\beta$-functions to one-loop order in the six-dimensional UED model.

\section{\label{sec:UED}Universal Extra Dimensions}
The extra dimensions in the UED model are flat and compactified. All SM fields are allowed to propagate through these extra dimensions. From our four-dimensional point of view they are perceived as an infinite tower of heavy KK modes corresponding to each SM particle, which, in turn, is identified with the zero-mode component. We consider a six-dimensional minimal UED model, with two extra, spatial dimensions, compactified on a so-called chiral square of side length $L$, which has the adjacent sides identified. A compactification radius $R=L/\pi$ is defined, and in this work, we will use the value $\mu_0=R^{-1}=1$ TeV, which is within the bounds on the size of the extra dimensions \cite{Belanger:2012mc}. The chiral square is equivalent to the $T^2/Z_4$ orbifold. We obtain a new conserved parity called KK parity. It is defined as $(-1)^{j+k}$ for the KK level with the KK numbers $j$ and $k$ and SM particles have KK numbers $j=0$ and $k=0$ \cite{Burdman:2005sr}.

Six-dimensional fermions have chiralities which are denoted $\pm$. The compactification is chosen so that the six-dimensional fermionic fields with $+$ chirality have left-handed zero-mode components \cite{Burdman:2006gy}. Hence, corresponding to a six-dimensional fermionic field with $+$ chirality, there is a tower of KK modes with chirality $+\ul$, which has a zero-mode component, and also a tower of chirality $+\ur$, without a zero-mode component. For the right-handed SM fermions, the expansions are given by interchanging $\ul$ and $\ur$ as well as $+$ and $-$. Furthermore, corresponding to each six-dimensional gauge field is a four-dimensional gauge field with a zero-mode component and two towers of real scalars, called adjoint scalars. By a suitable choice of expansion, these scalars will not have zero-mode components, but they do appear at the excited KK levels. The new KK particles interact with the SM particles, and hence, we obtain additional vertices involving the new fermionic fields and the adjoint scalars.

\begin{figure}[ht]
\begin{subfigure}
\centering
   \includegraphics[scale=0.7] {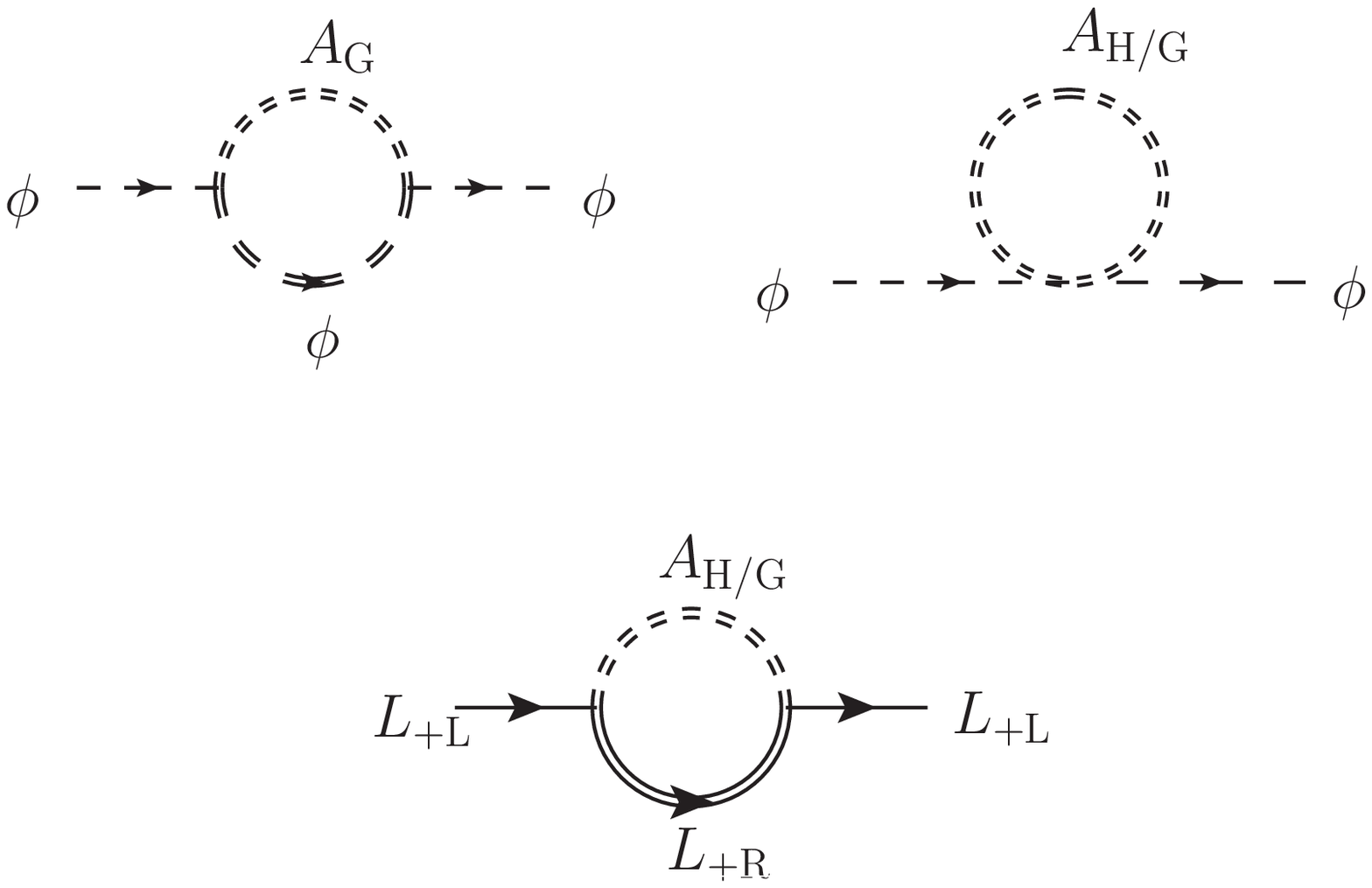}
   \caption*{\small{The new one-loop contributions to the corrections to the Higgs self-energy and the left-handed lepton doublet in the six-dimensional UED model. Here $A=B,W^i$, $i=1,2,3$.}}
   \label{fig:dia1}
 \end{subfigure}
 \begin{subfigure}
  \centering
   \includegraphics[scale=0.7] {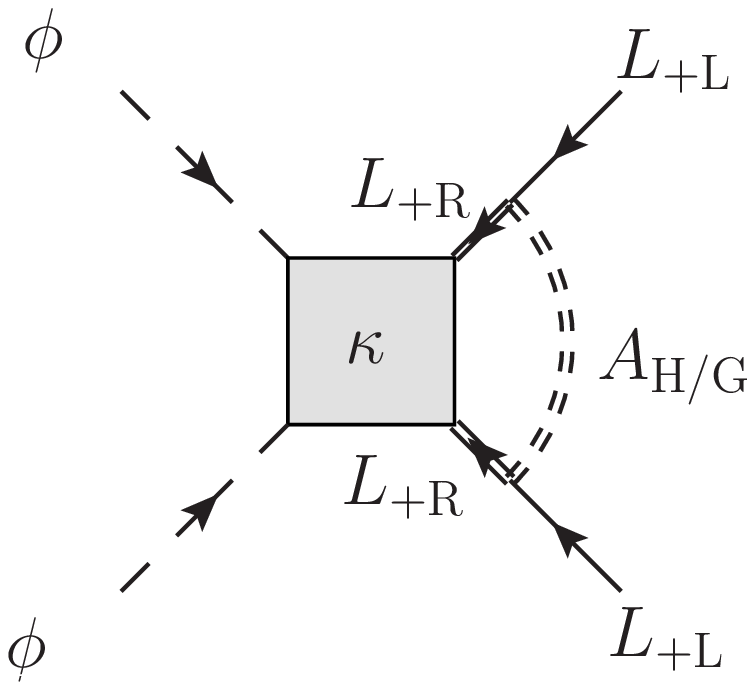}
   \caption*{\small{The new one-loop contribution to the correction of the dimensions-five Weinberg operator. Here $A=B,W^i$, $i=1,2,3$.}}
   \label{fig:dia2}
 \end{subfigure}
 \caption{The new Feynman diagrams, at one-loop order, containing the adjoint scalars and the fermion modes present at KK level, which contributes to the RG running of the neutrino mass parameters.}
 \label{fig:newdiagrams}
\end{figure}

In order to obtain neutrino masses in this model, we introduce a dimension-five Weinberg operator,
where two Higgs doublets are combined as an isospin triplet \cite{Weinberg:1980bf}. This is given by
\begin{eqnarray}\label{effectiveTerm}
\mathcal{L}_{\kappa}=-\frac{1}{8}\hat{\kappa}_{gf}(\overline{L_{\ul}^C}^g\varepsilon\tau^i L_{\ul}^f)(\phi^T\varepsilon\tau^i\phi)+\hc,
\end{eqnarray}
where $\kappa$ is a symmetric and complex matrix of mass dimension $-1$ in flavor space, $L$ denotes the lepton doublet fields, and $\phi$ the Higgs doublet fields. This operator can be generated through seesaw mechanisms. After electroweak symmetry breaking, the neutrino mass matrix is given by 
\begin{eqnarray}
m_{\nu}\equiv \kappa v^2, 
\end{eqnarray}
where $\kappa=\hat{\kappa}/L^2$ and $v\approx 174$ GeV is the Higgs vacuum expectation value. In UED models, the $\beta$-function for the neutrino mass matrix can be written on the form
\begin{eqnarray}
\frac{\mathrm{d}\kappa}{\mathrm{d}\ln \mu}=\beta^{\mathrm{SM}}_{\kappa}+s\beta^{\mathrm{UED}}_{\kappa},
\end{eqnarray}
where $\beta^{\mathrm{SM}}_{\kappa}$ is the SM contribution to the $\beta$-function and $\beta^{\mathrm{UED}}_{\kappa}$ are the contributions from the excited KK modes. The parameter $s$ counts the number of KK levels contributing to the total $\beta$-function, \ie ~the number of KK levels that fulfills $\sqrt{j^2+k^2}\leq\frac{\mu}{\mu_0}$, where $\mu$ is a given energy and $\mu_0=R^{-1}$ is the energy for which the first KK mode is excited. The contribution $\beta^{\mathrm{SM}}_{\kappa}$ has been calculated elsewhere, see \eg ~Refs.~\cite{Babu:1993qv,Chankowski:1993tx,Antusch:2001vn}, and it is given by
\begin{eqnarray}\label{eq:betaKSM}
\beta_{\kappa}^{\mathrm{SM}}&=&\frac{1}{16\pi^2}\left[2T\kappa+\lambda\kappa-\frac{3}{2}\kappa(Y_e^{\dagger}Y_e^{\phantom\dagger})-\frac{3}{2}(Y_e^{\dagger}Y_e^{\phantom\dagger})^{T}\kappa-3g_2^2\kappa\right]
\end{eqnarray}
with
\begin{eqnarray}
T=\mbox{Tr}\left(Y_e^{\dagger}Y_e^{\phantom\dagger}+3Y_u^{\dagger}Y_u^{\phantom\dagger}+3Y_d^{\dagger}Y_d^{\phantom\dagger}\right),
\end{eqnarray}
where the matrices $Y_f$ $(f=e,u,d)$ are the Yukawa couplings for the charged leptons, the up-type quarks, and the down-type quarks. Furthermore, the quantities $g_i$ $(i=1,2,3)$ are the gauge couplings and $\lambda$ is the Higgs self-coupling constant. The second contribution to the $\beta$-function, $\beta^{\mathrm{UED}}_{\kappa}$, comes from the excited KK modes. This contribution has been calculated in the six-dimensional UED model with the following result 
\begin{eqnarray}\label{eq:betaKUED}
\beta_{\kappa}^{\mathrm{UED}}&=&\frac{1}{16\pi^2}\left[4T\kappa+\lambda\kappa-\frac{3}{2}\kappa(Y_e^{\dagger}Y_e^{\phantom\dagger})-\frac{3}{2}(Y_e^{\dagger}Y_e^{\phantom\dagger})^{T}\kappa\right.-\left.\frac{3}{2}g_1^2\kappa-\frac{5}{2}g_2^2\kappa\right].
\end{eqnarray}
There is a difference between Eqs.~\eqref{eq:betaKSM} and \eqref{eq:betaKUED}, which is due to the new Feynman diagrams involving the new fermions and the adjoint scalars. The difference in the coefficients in front of the gauge couplings squared is due to the diagrams in Fig.~\ref{fig:newdiagrams}, whereas the factor of 2 in front of $T$ is based on new Yukawa-type interactions involving the new fermions at excited KK levels. The $\beta$-functions for the Yukawa matrices are listed in Appendix~\ref{sec:app}.

\section{\label{sec:RGrunning}Renormalization Group Running in Extra Dimensions}
The physical parameters are measured in low-energy scale experiments. Therefore, we need to take into account the effects of RG running when studying the parameters at high-energy scales \cite{Chankowski:1993tx,Antusch:2001ck,Antusch:2001vn,Blennow:2011mp}. All models with $d>4$ dimensions are inherently non-renormalizable. However, the theory can preserve its renormalizability if an energy cutoff, $\Lambda$, is introduced \cite{Dienes:1998vg}. Below $\Lambda$, the model is renormalizable and there we can thus treat it as an effective four-dimensional model. The energy cutoff implies that only a finite number of KK modes, namely the ones with energies between $R^{-1}$, where the first mode is excited, and $\Lambda$, contributes to the renormalization group (RG) running. Above $\Lambda$, an unknown, renormalizable ultraviolet completion is assumed to step in. Hence, the extra dimensions will affect the RG running of the physical parameters in the model. Instead of the logarithmic running that we have in the SM, the extra dimensions give rise to a power-law behavior of the RG running \cite{Dienes:1998vg}. Thus, it ought to be possible to distinguish the RG running in the UED model from that in the SM in the near future.

The contributions from the SM particles and the KK modes to the $\beta$-functions are different and independent. Therefore, they can be described in two separate terms and the RGEs for a quantity $Q$ is given by
\begin{eqnarray}
\frac{\rm{d}Q}{\rm{d}\ln \mu}=\beta_Q^{\mathrm{SM}}+s\beta_Q^{\mathrm{UED}},
\end{eqnarray}
where $\beta^{\mathrm{SM}}$ is the SM $\beta$-function and $\beta^{\mathrm{UED}}$ is the contribution from the KK modes at each KK level. We have assumed that the particle contents at each KK level are identical, except for the particle masses, and therefore that their contributions to the $\beta$-function are the same. The parameter $s$ counts the number of KK levels contributing to the total $\beta$-function. Below $\mu_0$, where the first KK mode is excited, is the running logarithmic, since only the SM particles contribute to the $\beta$-function. As the energy increases from $\mu_0$ to $\Lambda$, the energy threshold for successive modes are crossed, and thus, an increasing number of KK modes comes into play. For energies fulfilling $\mu\gg\mu_0$, the contributions from the KK modes will dominate and the RG running will have a power-law dependence.
\section{\label{sec:RGHiggs}Running of the Higgs Self-Coupling}
First, we investigate the bounds on the Higgs self-coupling from the vacuum stability criterion. Bounds on the five- and six-dimensional UED models from LHC data have previously been studied in Refs.~\cite{Nishiwaki:2011gk,Blennow:2011tb,Nishiwaki:2011gm,Cornell:2011ge,Belanger:2012mc}. We have calculated the RGEs for the Higgs self-coupling, $\lambda$, in the six-dimensional UED model and they are presented in Eqs.~\eqref{eq:higgsrunning}--\eqref{eq:betaL}. The RG running of $\lambda$ can be used to put constraints on this model. Depending on the initial conditions, there are two possible scenarios, which limit the stability region of $\lambda$. Either it approaches the triviality limit, where it diverges, or the vacuum stability limit, where the value of $\lambda$ becomes negative and the Higgs potential becomes unstable. For the Higgs mass, we use the mass recently determined by the ATLAS and CMS Collaborations, $m_{\rm H}=126\GeV$ \cite{:2012gk,:2012gu}. This mass is in the low-mass region, and therefore only the triviality limit is an issue.

\begin{figure}
\includegraphics[scale=0.5]{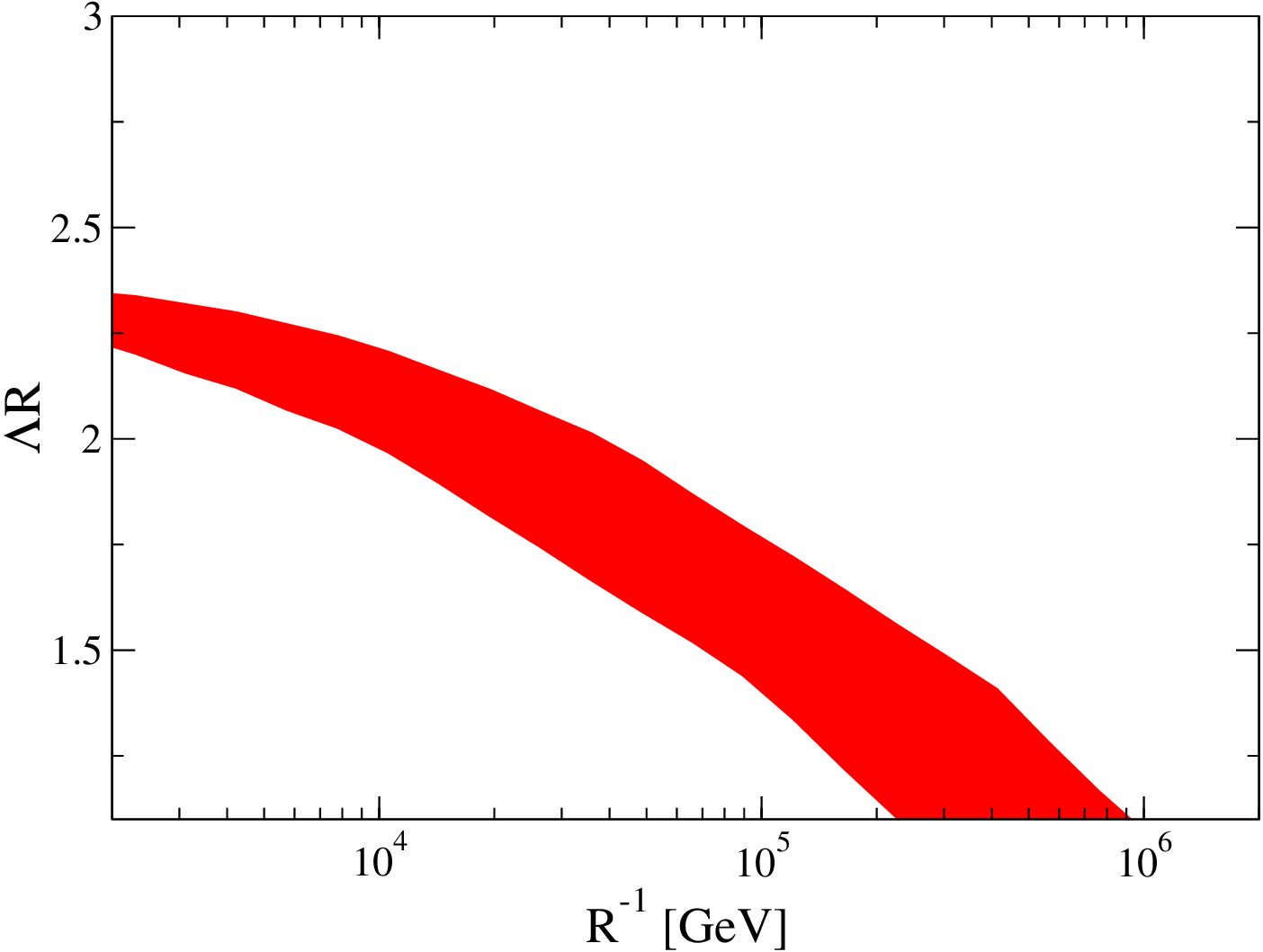}
\caption{The upper bounds on $\Lambda R$ as a function of $R^{-1}$ for a Higgs mass of $m_{\rm H}=126\GeV$ from the vacuum stability criterion for the Higgs self-coupling constant. The band is due to the uncertainty in the top quark mass, which is in the range $170.9\GeV-173.3\GeV$, and the strongest bound is due to the largest mass.}
\label{fig:higgs}
\end{figure}

We have solved the full set of RGEs numerically, and in Fig.~\ref{fig:higgs}, we present the bounds on $\Lambda R$ as a function of $R^{-1}$ from the vacuum stability criterion of the Higgs self-coupling constant. The product $\Lambda R$ counts the number of KK modes below the cutoff scale, \ie ~the number contributing to the RGEs. The mass of the top quark, $m_t$, is only known with a precision of a few GeV and it affects the bounds on the model. Here we have used $170.9\GeV-173.3\GeV$ \cite{Xing:2011aa}. The weakest limit on the number of KK modes contributing is given at 1 TeV and even there only five modes are in the energy range and thus can contribute, \ie ~the modes with KK numbers $(j,k)=(1,0),(1,1),(2,0),(1,2),(2,1)$, which is a severe constraint on the validity of the extra-dimensional model. This result is in line with the five-dimensional case, which has been investigated in Ref.~\cite{Blennow:2011tb}, where only five KK modes were allowed from the vacuum stability criterion. Note that our results are based on the $\beta$-functions calculated to one-loop order. Hence, there may be small changes due to higher-order corrections.

\section{\label{sec:RGMixing}Running of the Leptonic Mixing Angles}
We will proceed our discussion to the RG running of the neutrino parameters, and more specifically to the running of the leptonic mixing angles. In order to determine the RG running of the mixing angles, we use the standard parametrization, where the mixing matrix, $U$, is parametrized in terms of three mixing angles, $\theta_{12}$, $\theta_{13}$, and $\theta_{23}$ and three CP-violating phases, $\delta$, $\rho$, and $\sigma$. The value of $\theta_{13}$ is non-zero in contrast to what has been previously expected. In fact, it has a comparatively large value $\sin^2(\theta_{13})=0.023\pm0.0023$ at $1\sigma$ confidence level \cite{Schwetz:2012}. A non-zero value of $\theta_{13}$ strongly disfavors the bimaximal and tri-bimaximal mixing patterns. Several global fits have been performed to the experimental results from neutrino oscillation experiments \cite{Tortola:2012te,Fogli:2012ua}. We have taken the values of the neutrino mixing parameters at $M_Z$, being the $Z$ boson mass, from Ref.~\cite{Schwetz:2012}. It should however be pointed out that our results are rather independent of the global fit used.

The standard parametrization of the mixing matrix is given by
\begin{eqnarray}
U=\left( \begin{array}{ccc}
c_{12}c_{13} & s_{12}c_{13} & s_{13}{\rm e}^{-\im\delta} \\
-s_{12}c_{23}-c_{12}s_{23}s_{13}{\rm e}^{\im\delta} & c_{12}c_{23}-s_{12}s_{23}s_{13}{\rm e}^{\im\delta} & s_{23}c_{13} \\
s_{12}s_{23}-c_{12}c_{23}s_{13}{\rm e}^{\im\delta} & -c_{12}s_{23}-s_{12}c_{23}s_{13}{\rm e}^{\im\delta} & c_{23}c_{13} \end{array} \right)\left( \begin{array}{ccc}
{\rm e}^{\im\rho} & 0 & 0 \\
0 & {\rm e}^{\im\sigma}& 0\\
0 & 0& 1 \end{array} \right),
\end{eqnarray}
where $s_{ij}=\sin(\theta_{ij})$ and $c_{ij}=\cos(\theta_{ij})$, $ij=12,13,23$. In the present work, we will assume $\delta=\rho=\sigma=0$.

The neutrino mass matrix can then be diagonalized by
\begin{eqnarray}
U^{\dagger}m_{\nu}U^*=\mathrm{diag}(m_1,m_2,m_3),
\end{eqnarray}
where $m_i$, $(i=1,2,3)$ are the neutrino masses. There are two possible orderings of the neutrino masses, which are in agreement with experimental data. Normal hierarchy, where $m_1<m_2<m_3$, or inverted hierarchy, where $m_3<m_1<m_2$. The neutrino mass scale is yet unknown, but there are upper bounds on the neutrino masses coming from tritium beta decay experiments, \ie~$m_{\nu}<2$ eV \cite{Otten:2008zz}. These bounds will most probably be improved by one order of magnitude by the KATRIN experiment \cite{Sturm:2011ms,Osipowicz:2001sq}. Apart from direct bounds, there are also indirect constraints from cosmological data, such as the CMB data of the WMAP experiment, which are more severe and put an upper limit on the sum of the neutrino masses \cite{Komatsu:2010fb}. In the present work, we will assume $m_{\nu}\approx 0.5$ eV.

In the numerical analysis, the full set of RGEs have been used, these are given in Appendix \ref{sec:app}. In Fig.~\ref{fig:angles}, we display the running of the neutrino mixing angles as functions of the energy in the cases of normal and inverted hierarchy. The mixing angle $\theta_{12}$ has the most significant RG dependence, in both normal and inverted hierarchies, and in accordance with other extra-dimensional models, the RG running exhibits a power-law behavior. Neither $\theta_{13}$ nor $\theta_{23}$ exhibits any significant RG running in either case. It should, however, be noted that the direction of the small running depends on the mass hierarchy for both $\theta_{13}$ and $\theta_{23}$. The value increases (decreases) in case of normal (inverted) hierarchy. In both hierarchies, it is possible that $\theta_{12}=\theta_{23}$ close to the cutoff scale if $\theta_{12}$ is large and $\theta_{23}$ is small at $M_Z$, which would imply a bimaximal mixing pattern, however with a non-zero $\theta_{13}$. The RG running of the mixing angles in the five-dimensional UED model was discussed in Ref.~\cite{Blennow:2011mp}. The RG running is similar in both cases. The only mixing angle that exhibits any significant RG running is $\theta_{12}$, and furthermore, $\theta_{13}$ and $\theta_{23}$ run in the same direction for the respective hierarchies.  It turns out that the main difference between the two models is the cutoff scale, which in the five-dimensional model is taken to be $\Lambda=50$ TeV, which is significantly larger than in the six-dimensional case, where $\Lambda=6.5$ TeV. Hence, the six-dimensional model ought to be detectable in a closer future. Furthermore, the low value of $\lambda$ means that the energy required for unification of the gauge parameters cannot be attained.
\begin{figure}[ht]
\subfigure{
   \includegraphics[scale =0.3,clip] {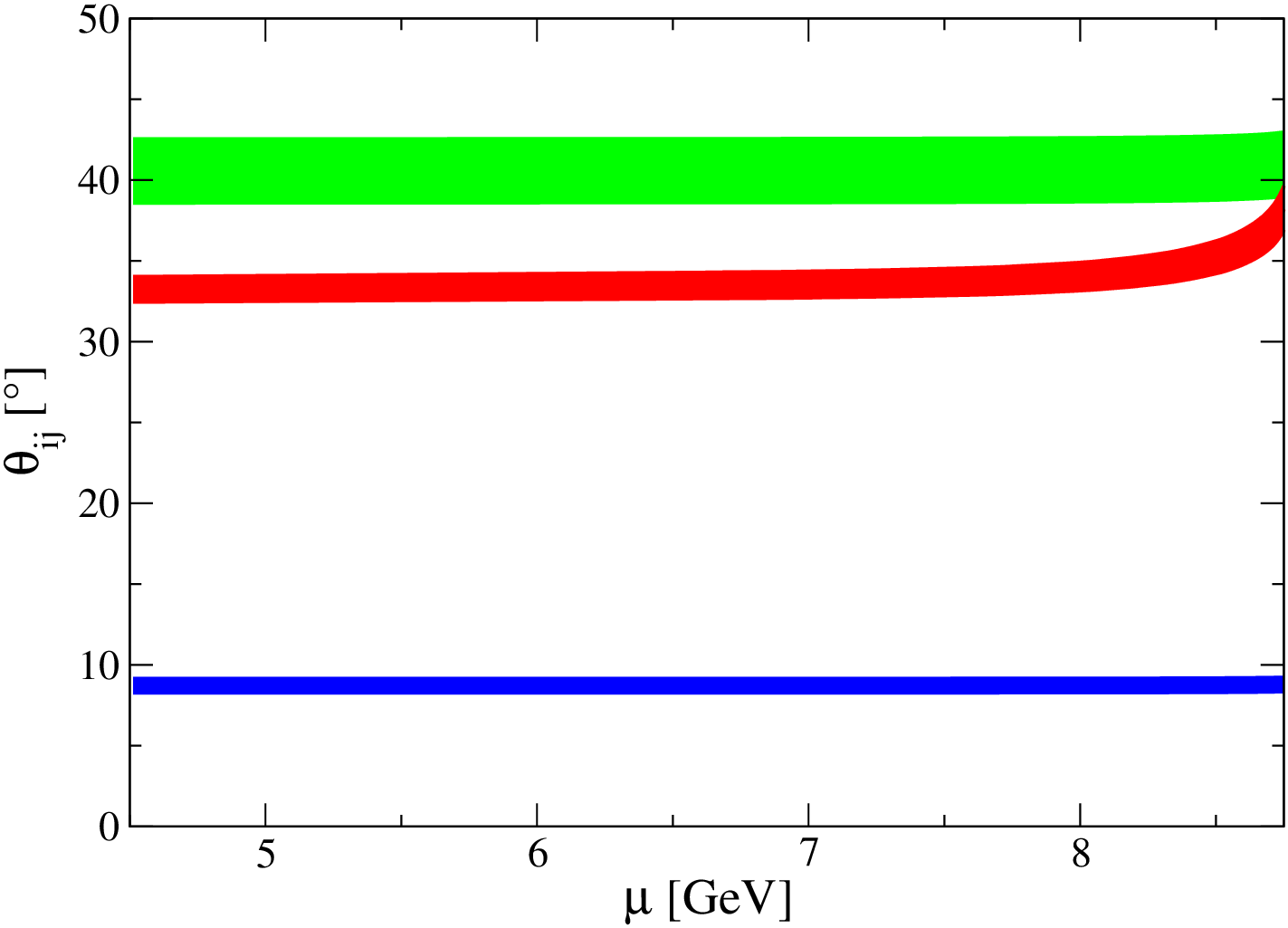}
   \label{fig:subfig1}
 }
 \subfigure{
   \includegraphics[scale =0.3,clip] {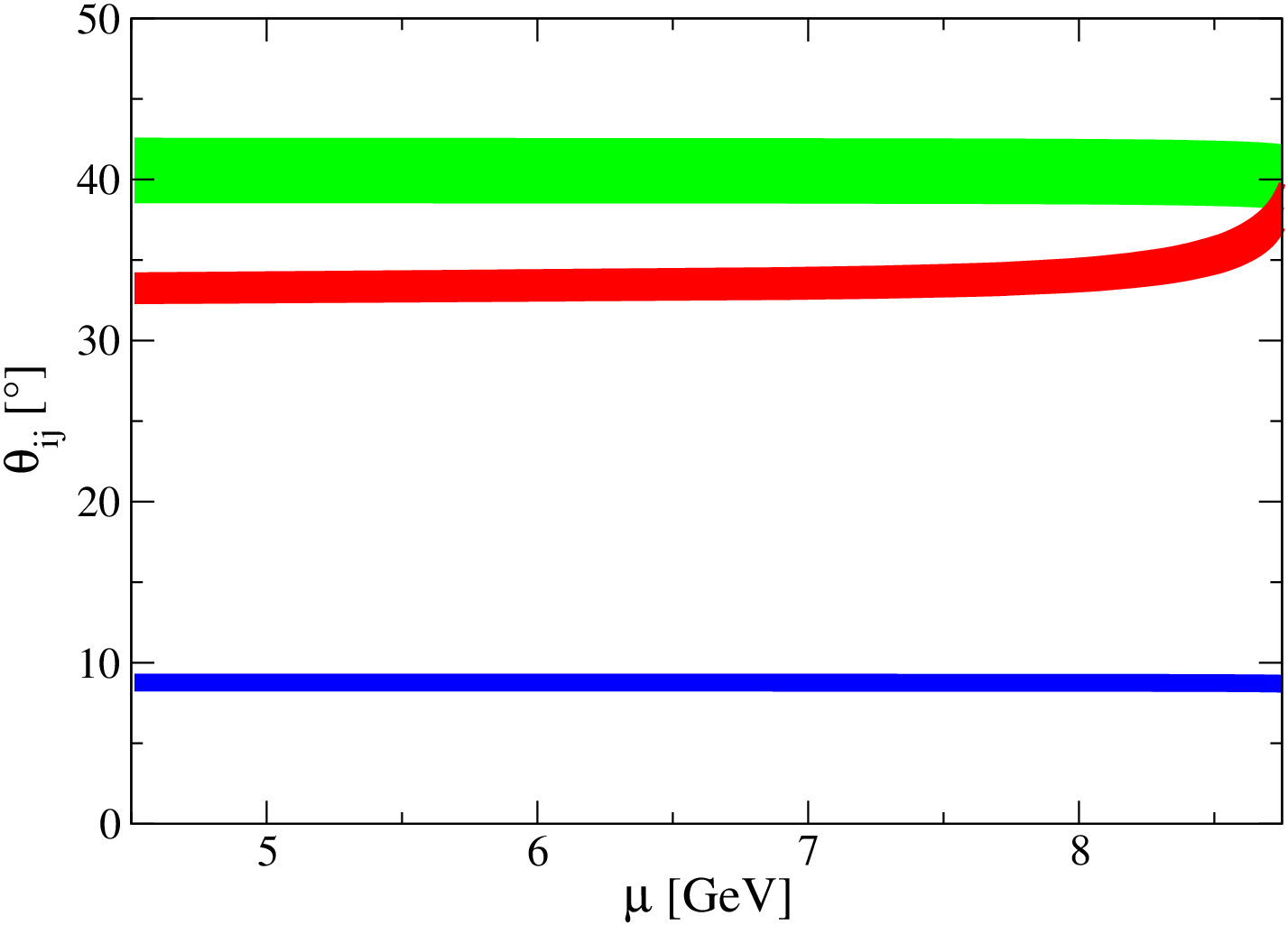}
   \label{fig:subfig2}
 }
 \caption{The RG evolution of the three leptonic mixing angles as functions of the energy scale. The runnings are from $M_Z$ to the cut-off scale, $\Lambda$, for the case of normal hierarchy (left plot) and inverted hierarchy (right plot). The different bands correspond to the respective $1\sigma$ confidence intervals. The bands corresponding to the $2\sigma$ and $3\sigma$ confidence intervals are significantly wider, which we have checked numerically but not presented here. The lightest neutrino mass is assumed to be 0.5 eV.} \label{fig:angles}
\end{figure}

The cut-off energy in the present six-dimensional UED model is comparatively low, and therefore, neutrino mass operators of mass dimension $d>5$ might change the RG running of the neutrino masses. A classification of such operators has been presented in Ref.~\cite{Babu:2001ex} and they can be generated in so-called radiative models. At one-loop order, \ie~of mass dimension $d=7$, there are seven such operators \eg
\begin{eqnarray}
\mathcal{O}=L^iL^jL^ke^cH^l\epsilon_{ij}\epsilon_{kl}.
\end{eqnarray}
Certainly, such operators will change the $\beta$-function for the neutrino mass operator, and could indeed give rise to both qualitative and quantitative differences in the RG running compared to the situation without including any higher-order mass operators. Thus, by adding more mass operators, we would change the original model, and such a study lies beyond the scope of this work. Note that other studies of higher-order mass operators have been made in e.g. Refs.~\cite{Krauss:2011ur,Bonnet:2009ej,Bonnet:2012kz,DeGouvea:2001mz}. To our knowledge, no study of the effects of higher-order mass operators in the context of UEDs has yet been performed.

\section{\label{sec:summary}Summary and conclusions}
In this work, we have studied the RG running of physical parameters in a six-dimensional UED model, extended by a dimension-five effective neutrino mass operator. We have specifically derived the RGEs for the Yukawa couplings, the effective neutrino mass operator, and the Higgs self-coupling in this model.

Furthermore, we have studied the RG evolution of the leptonic mixing angles numerically, and found that the angle $\theta_{12}$ shows the most sizable running and is always increasing, independent of the mass ordering. The RG corrections to the mixing angles $\theta_{13}$ and $\theta_{23}$ are negligible. It can, however, be noted that the direction of the running is in their case dependent on the mass hierarchy. The values of both mixing angles  are increasing, albeit only a little, for normal ordering, and decreasing for inverted ordering.

Finally, we have used the new value of the Higgs mass determined at the LHC and the vacuum stability criterion to set a bound on the extra-dimensional model. It turns out that only five KK modes are allowed to contribute, which severely limits the validity of the UED model and puts a severe limitation on the value of the cut-off as well. In particular, it means that the energy required for unification of the gauge couplings is not reached within this model.
\section*{\label{sec:ack}Acknowledgments}
We would like to thank Henrik Melb{\'e}us for useful and stimulating discussions in the initial phase of this work, He Zhang for information about and help with the numerics, and Shun Zhou for useful discussions. This work was supported by the Swedish Research Council (Vetenskapsr{\aa}det), contract No.~621-2011-3985 (T.O.).
\appendix
\section{\label{sec:app} Full Set of RGEs to One-loop Order in the Six-dimensional UED model}
The running of the Higgs self-coupling is given by
\begin{eqnarray}\label{eq:higgsrunning}
\beta_{\lambda}=\beta_{\lambda}^{\mathrm{SM}}+s\beta_{\lambda}^{\mathrm{UED}}.
\end{eqnarray}
The SM contribution to the $\beta$-function is given in Ref.~\cite{Cheng:1973nv}
\begin{eqnarray}\label{eq:HiggsSM}
\beta_{\lambda}^{\mathrm{SM}}&=&\frac{1}{16\pi^2}\Big\{12\lambda^2+\frac{3}{4}(g_1^4+2g_1^2g_2^2+3g_2^4)-3\lambda(g_1^2+3g_2^2)\notag\\&\phantom+&+4\lambda T-4\mbox{Tr}\left[(Y_e^{\dagger}Y_e^{\phantom\dagger})^2+3(Y_u^{\dagger}Y_u^{\phantom\dagger})^2+3(Y_d^{\dagger}Y_d^{\phantom\dagger})^2\right]\Big\}.
\end{eqnarray}
where
\begin{eqnarray}\label{eq:T} T=\mbox{Tr}\left(Y_e^{\dagger}Y_e^{\phantom\dagger}+3Y_u^{\dagger}Y_u^{\phantom\dagger}+3Y_d^{\dagger}Y_d^{\phantom\dagger}\right)
\end{eqnarray}
The UED contribution to the $\beta$-function is determined to be
\begin{eqnarray}\label{eq:betaL}
\beta_{\lambda}^{\mathrm{UED}}&=&\frac{1}{16\pi^2}\Big\{12\lambda^2+\frac{5}{4}(g_1^4+2g_1^2g_2^2+3g_2^4)-3\lambda(g_1^2+3g_2^2)\notag\\ &+& 8\lambda T- 8\mbox{Tr}\left[(Y_e^{\dagger}Y_e^{\phantom\dagger})^2+3(Y_u^{\dagger}Y_u^{\phantom\dagger})^2+3(Y_d^{\dagger}Y_d^{\phantom\dagger})^2\right]\Big\}.
\end{eqnarray}
\\
The running of the Yukawa couplings are given by 
\begin{eqnarray}
\beta_{Y_{i}}=\beta_{Y_{i}}^{\mathrm{SM}}+s\beta_{Y_{i}}^{\mathrm{UED}},
\end{eqnarray}
where $i=e,u,d$. The SM contributions to the $\beta$-functions for the Yukawa couplings can be found in, \eg, Ref.~\cite{Grzadkowski:1987wr} and they are given by
\begin{eqnarray}
\beta^{\mathrm{SM}}_{Y_e}&=&\frac{1}{16\pi^2}Y_e^{\phantom\dagger}\left[T+\frac{3}{2}Y_e^{\dagger}Y_e^{\phantom\dagger}-\frac{15}{4}g_1^2-\frac{9}{4}g_2^2\right],\\
\beta^{\mathrm{SM}}_{Y_u}&=&\frac{1}{16\pi^2}Y_u^{\phantom\dagger}\left[T+\frac{3}{2}Y_u^{\dagger}Y_u^{\phantom\dagger}-\frac{3}{2}Y_d^{\dagger}Y_d^{\phantom\dagger}-\frac{17}{12}g_1^2-\frac{9}{4}g_2^2-8g_3^2\right],\\
\beta^{\mathrm{SM}}_{Y_d}&=&\frac{1}{16\pi^2}Y_d^{\phantom\dagger}\left[T+\frac{3}{2}Y_d^{\dagger}Y_d^{\phantom\dagger}-\frac{3}{2}Y_u^{\dagger}Y_u^{\phantom\dagger}-\frac{17}{12}g_1^2-\frac{9}{4}g_2^2-8g_3^2\right].
\end{eqnarray}
The UED contributions to the $\beta$-functions are given by 
\begin{eqnarray}\label{eq:Y1}
\beta^{\mathrm{UED}}_{Y_e}&=&\frac{1}{16\pi^2}Y_e^{\phantom\dagger}\left[2T+\frac{3}{2}Y_e^{\dagger}Y_e^{\phantom\dagger}-\frac{1}{2}g_1^2-\frac{3}{2}g_2^2\right],\\
\beta^{\mathrm{UED}}_{Y_u}&=&\frac{1}{16\pi^2}Y_u^{\phantom\dagger}\left[2T+\frac{3}{2}Y_u^{\dagger}Y_u^{\phantom\dagger}-\frac{3}{2}Y_d^{\dagger}Y_d^{\phantom\dagger}-\frac{1}{2}g_1^2-\frac{3}{2}g_2^2\right],\\
\beta^{\mathrm{UED}}_{Y_d}&=&\frac{1}{16\pi^2}Y_d^{\phantom\dagger}\left[2T+\frac{3}{2}Y_d^{\dagger}Y_d^{\phantom\dagger}-\frac{3}{2}Y_u^{\dagger}Y_u^{\phantom\dagger}-\frac{1}{2}g_1^2-\frac{3}{2}g_2^2\right]\label{eq:Y2}.
\end{eqnarray}
The SM contribution to the $\beta$-function for the effective neutrino mass operator is given by
\begin{eqnarray}
\beta_{\kappa}^{\mathrm{SM}}&=&\frac{1}{16\pi^2}\left[2T\kappa+\lambda\kappa-\frac{3}{2}\kappa(Y_e^{\dagger}Y_e^{\phantom\dagger})-\frac{3}{2}(Y_e^{\dagger}Y_e^{\phantom\dagger})^{T}\kappa-3g_2^2\kappa\right].
\end{eqnarray}
The contribution from the extra dimensions is given by
\begin{eqnarray}
\beta_{\kappa}^{\mathrm{UED}}&=&\frac{1}{16\pi^2}\left[4T\kappa+\lambda\kappa-\frac{3}{2}\kappa(Y_e^{\dagger}Y_e^{\phantom\dagger})-\frac{3}{2}(Y_e^{\dagger}Y_e^{\phantom\dagger})^{T}\kappa\right.-\left.\frac{3}{2}g_1^2\kappa-\frac{5}{2}g_2^2\kappa\right].
\end{eqnarray}\\
Finally, we state the RGEs for the gauge couplings which have been calculated elsewhere, \eg ~Ref.~\cite{Nishiwaki:2011gm}, but are presented here for the sake of completeness
\begin{eqnarray}
16\pi^2\dd{g_i}{\ln \mu}=(b_k+s\bar{b}_k)g_i^2,
\end{eqnarray} 
where $b_1=41/6,b_2=-19/6,b_3=-7$ and $\bar{b}_1=27/2,\bar{b}_2=3/2, \bar{b}=-2$.


\end{document}